\documentclass[mathleft, cite]{an}
\usepackage{graphicx, psfig}
\usepackage{amsfonts}
\usepackage{times}
\begin{document}
\Pagespan{1}{}%
\Yearpublication{2006}%
\Yearsubmission{2006}%
\Month{9}%
\Volume{999}%
\Issue{88}%
\DOI{DOI}%

\title{Estimation of Galactic model parameters with the Sloan Digital Sky Survey and metallicity distribution in two fields in the anti-centre direction of the Galaxy}

\author{S. Ak\inst{1}\fnmsep\thanks{Corresponding author:\email{akserap@istanbul.edu.tr}\newline} 
\and  S. Bilir\inst{1}
\and  S. Karaali\inst{2}
\and  R. Buser\inst{3}
}
\institute{Istanbul University Science Faculty, Department of Astronomy and Space Sciences, 34119, University-Istanbul, Turkey
\and
Beykent University, Faculty of Science, Department of Mathematics and Computer, Beykent 34398, Istanbul, Turkey
\and
Astronomisches Institut der Universit\"{a}t Basel, Venusstrasse 7, 4102 Binningen-Switzerland}
\date{} 
\received{}
\accepted{}
\publonline{later} 


\abstract{We estimated the Galactic model parameters from the Sloan Digital Sky Survey ({\em SDSS}) data reduced for two fields in the anti-centre direction of the Galaxy, $l= 180^{o}$, and symmetric relative to the Galactic plane, $b=+45^{o}$ (north field) and $b=-45^{o}$ (south field). The large size of each field, 60 deg$^{2}$, and the faint limiting apparent magnitude, $g_{0}=22$, give us the chance to determine reliable parameters for three components, thin and thick discs and halo, in the north and south hemispheres of the Galaxy, except the scalelengths for two discs which are adopted from Juri\'c et al. (\cite{J05}). Metallicities were evaluated by a recent calibration for {\em SDSS}, and absolute magnitudes of stars with $4<M(g)\leq8$ were derived as a function of $(g-r)_{0}$ colour and metallicity. A $\chi^{2}$ method was employed to fit the analytical density laws to the observational-based space densities with the addition constraint of producing local densities consistent with those derived from Hipparcos. Conspicuous differences could not be detected between the corresponding Galactic model parameters for the thin disc of north and south fields, and our results are consistent with the ones in the literature. The same case is valid for the halo, especially the axis ratios for two fields are exactly equal, $\kappa=0.45$, and close to the one of Juri\'c et al. (2005). However, we revealed differences between the scaleheights and local space densities for the thick disc of the north and south fields. The metallicity distribution for unevolved G type stars with $5<M(g)\leq6$ shows three substructures relative to the distance from the Galactic plane: for $z^{*}<3$ kpc, the metallicity gradient for two fields is $d[M/H]/dz \sim -0.20(\pm0.02)$ dex kpc$^{-1}$, consistent with the formation scenario of the thin disc. For $5<z^{*}<10$ kpc, $d[M/H]/dz \sim -0.03(\pm0.001)$ dex kpc$^{-1}$ for two fields, confirming that the metallicity gradient for the halo component is close to zero. However, the tendency of the metallicity for stars with $3<z^{*}<5$ kpc, corresponding to the transition region from thick disc to halo is different. For the north field $d[M/H]/dz=-0.36(\pm0.12)$ dex kpc$^{-1}$, whereas it is half of this value for the south field, $d[M/H]/dz=-0.18(\pm0.01)$ dex kpc$^{-1}$. The origin of this conflict is probably due to the different structure of the thick disc in opposite latitudes of the Galaxy. When we combine these substructures, however, we find a smooth metallicity gradient for two fields, $-0.30(\pm0.04)$ dex kpc$^{-1}$.     
}

\maketitle
\section{Introduction}
The study of Galactic models and parameters has a long history. Bahcall \& Soneira (\cite{BS80}) fitted observations with a two-component Galactic model, namely disc and halo, while Gilmore \& Reid (\cite{GR83}) success was to fit their observations with a Galactic model introducing a third component, i.e. the thick disc. It should be noted that the third component was a rediscovery of the ``Intermediate Population II" first described in the Vatican Proceedings review of O'Connell (\cite{O58}). Due to their importance, Galactic models have been attractive for many research centers: they can be used as a tool for uncovering the formation and evolution of the Galaxy.

Different research groups have been using different methods to determine parameter values, and these have continued to cover large ranges through the
most recent work (Table 1). For example, Chen et al. (\cite{C01}) and Siegel et al. (\cite{S02}) give 6.5--13 and 6--10 per cent, respectively, for the relative local space density of the thick disc. However, we feel that an appropriate procedure would lead us to expect a smaller range and/or a unique value carrying a small error. Indeed, in previous papers (Karaali, Bilir \& Hamzao\u glu \cite{KBH04}, hereafter KBH; Bilir, Karaali \& Gilmore \cite{BKG06}, hereafter BKG) we argued that Galactic model parameters must be mass dependent and, since absolute magnitude provides a reasonable proxy for mass, they must vary from one absolute magnitude interval to the next. Therefore, neglect of this mass-or absolute magnitude-dependence leads to large errors in parameter values, and this probably also explains their large ranges exhibited by the star-count studies listed in Table 1.

Since of course, both the data and the parameter estimation method affect the quality of the model parameter values in a crucial way, the present project has been designed as follows.

Our data will consist of homogeneous imaging sets of {\em SDSS\/} for stars in several fields in different directions of the Galaxy. Using these data, we shall apply a parameter estimation procedure which is based on the most recent calibrations: 1) absolute magnitudes will be derived as functions of both $(g-r)_{0}$ colour and metallicity and 2) metallicities will be determined using the new calibration provided by Karaali, Bilir \& Tun\c{c}el (\cite{KBT05}, hereafter KBT).
 
The ultimate goal of this project is to derive, in a series of papers, more reliable values of Galactic model parameters, including their dependence on Galactic longitude and latitude.

In Section 2 we describe the {\em SDSS\/} data and their reductions, and in Section 3 we present the density laws that we shall adopt in the present study. Sections 4 and 5 provide the metallicity calibration, the absolute magnitude determination, and the evaluation of the density functions for individual population components. In Section 6 we estimate the stellar metallicities and final conclusions are drawn in Section 7.

\section{The Sloan Digital Sky Survey}
The {\em SDSS\/} is a large, international collaboration project set up to survey 10 000 square--degrees of sky in five optical passbands and to obtain spectra of one million galaxies, 100 000 quasars, and tens of thousands of Galactic stars. The data are being taken with a dedicated 2.5m telescope located at Apache Point Observatory (APO), New Mexico. The telescope has two instruments: a CCD camera with 30 2048$\times$2048 CCDs in the focal plane and two 320 fiber double spectrographs. The imaging data are tied to a network of brighter astrometric standards (which would be saturated in the main imaging data) through a set of 22 smaller CCDs in the focal plane of the imaging camera. An 0.5m telescope at APO will be used to tie the imaging data to brighter photometric standards. 

The {\em SDSS\/} obtains images almost simultaneously in five broad bands ($u$, $g$, $r$, $i$ and $z$) centered at 3540, 4760, 6280, 7690 and 9250 $\AA$, respectively (Fukugita et al. \cite{F96}, Gunn et al. \cite{G98}, Smith et al. \cite{Smith02}, Hogg et al. \cite{H02}). The imaging data are automatically processed through a series of software pipelines which find and measure objects and provide photometric and astrometric calibrations to produce a catalogue of objects with calibrated magnitudes, positions and structure information. The photometric pipeline (Lupton et al. \cite{L01}) detects the objects, matches the data from the five filters, and measures instrumental fluxes, positions and shape parameters. The last allow the classification of objects as ``point source'' (compatible with the point spread function), or ``extended''. The psf magnitudes are currently accurate to about 2 per cent in $g$, $r$ and $i$ and 3--5 per cent in $u$ and $z$ for bright ($<20^{m}$) point sources. The {\em SDSS\/} in Data Release 4 (DR4) is almost 85 per cent complete for point  sources to ($u$, $g$, $r$, $i$, $z$) = (22, 22.2, 22.2, 21.3, 20.5), and the full-width at half-maximum of the psf is about 1.5 arcsec (Abazajian et al. \cite{A04}). The data are saturated at about 14$^{m}$ in $g$, $r$ and $i$ and about 12$^{m}$ in $u$ and $z$. Astrometric calibration is carried out using a second set of less sensitive CCDs in the camera, which allows the transfer of astrometric catalogue positions to the fainter objects detected by the main camera. Absolute positions are accurate to better than 0.1 arcsec rms in each coordinate (Pier et al. \cite{P03}).

\subsection{Data and reductions}

The data are taken from {\em SDSS\/} (DR4) on the WEB\footnote {http://www.sdss.org/dr4/access/index.html} for two fields with the same Galactic longitude ($l=180^{o}$) and symmetric relative to the Galactic plane, $b=+45^{o}$ (hereafter north) and $b=-45^{o}$ (hereafter south). The areas of the fields are equal and rather large, 60 deg$^{2}$, avoiding any fluctuation in the values of Galactic model parameters. {\em SDSS\/} magnitudes $u$, $g$, $r$, $i$ and $z$ are available for 516 279 point sources in two fields down to the limiting magnitude of $g_{o}=24$. The total absorptions, $A_{u}$, $A_{g}$, $A_{r}$, $A_{i}$ and $A_{z}$, for each point source are taken from {\em SDSS\/} query server, and the apparent magnitudes $u$, $g$, $r$, $i$ and $z$ are de-reddened by equations in eq. (1)  

\setlength{\mathindent}{0pt}
\begin{eqnarray}
u_{0} = u - A_{u}, g_{0} = g - A_{g}, r_{0} = r - A_{r}\nonumber\\
i_{0} = i - A_{i}, z_{0} = z - A_{z},
\end{eqnarray}
All the colours and magnitudes mentioned hereafter will be de-reddened ones.

Star/galaxy separation was performed by utilizing the command prob {\em PSFmag} in DR4 WEB page, to provide each object's probability of being a star in each filter (parameter values of 0 or 1 indicating a galaxy or star, respectively). The quality of this separation therefore strongly depends on the seeing and sky brightness. In addition to the work cited above, we adopted the procedure of Juri\'c et al. (\cite{J05}) in order to remove the scattered point sources in the $(r-i)_{0}-(g-r)_{0}$ diagram. Thus, sources at distances larger than 0.3 mag from the locus were excluded from the statistics (Fig. 1). Such large deviations are inconsistent with measurement errors, and in most cases indicate a source that is not a main sequence star. This requirement effectively removes hot white dwarfs (Kleinman et al. \cite{K04}), low-redshift quasars ($z<2.2$, Richards et al. \cite{R01}), and white/red dwarf unresolved binaries (Smol\v ci\'c et al. \cite{S04}).

The apparent magnitude histograms for point sources and for the star sample are shown in Fig. 2, where the limiting apparent magnitude for the survey stars can be read off as $g_{0}=22^{m}$. Owing to the {\em SDSS} observing strategy, stars brighter than $g_{0}=14^{m}$ will be saturated, and star counts will be not complete for magnitudes fainter than $g_{0}=22^{m}.2$ (Abazajian et al. \cite{A04}). Hence our work is limited for the magnitude range $15^{m}<g_{0}\leq22^{m}$ for the evaluation of reliable Galactic model parameters.

\section{Density laws}

Disc structures are usually parametrized in cylindrical coordinates by radial and vertical exponentials,

\setlength{\mathindent}{0pt}
\begin{eqnarray}
D_{i}(x,z)=n_{i}~exp(-|z|/H_{i})~exp(-(x-R_{0})/h_{i}),
\end{eqnarray}
where $z=z_{\odot}+r\sin(b)$, $r$ is the distance to the object from the Sun, $b$ is the Galactic latitude, $z_{\odot}$ is the vertical distance of the Sun from the Galactic plane (24 pc, Juri\'c et al. \cite{J05}), $x$ is the planar distance from the Galactic centre, $R_{0}$ is the solar distance from the Galactic centre (8 kpc, Reid \cite{R93}), $H_{i}$ and $h_{i}$ are the scaleheight and scalelength, respectively, and $n_{i}$ is the normalized density at the solar radius. The suffix $i$ takes the values 1 and 2 as long as the thin and thick discs are considered. 

The density law for the spheroid component is parameterized in different forms. The most common is the de Vaucouleurs (\cite{deva48}) spheroid used to describe the surface brightness profile of elliptical galaxies. This profile has been deprojected into three dimensions by Young (\cite{Y76}) as 

\setlength{\mathindent}{0pt}
\begin{eqnarray}
D_{s}(R)=n_{s}~exp[-7.669(R/R_{e})^{1/4}]/(R/R_{e})^{7/8},
\end{eqnarray}
where $R$ is the (uncorrected) Galactocentric distance in spherical coordinates, $R_{e}$ is the effective radius and $n_{s}$ is the normalized local density. $R$ has to be corrected for the axial ratio $\kappa = c/a$, 

\setlength{\mathindent}{0pt}
\begin{eqnarray}
R = [x^{2}+(z/\kappa)^2]^{1/2},
\end{eqnarray}
where,
\setlength{\mathindent}{0pt}
\begin{eqnarray}
z = r \sin b,
\end{eqnarray}
\setlength{\mathindent}{0pt}
\begin{eqnarray}
x = [R_{0}^{2}+r^{2}\cos^{2} b-2R_{0}r\cos b \cos l]^{1/2}, 
\end{eqnarray}
$r$ being the distance along the line of sight and, $b$ and $l$ the Galactic latitude and longitude respectively, for the field under investigation.
The form used by the Basel group is independent of effective radius but is dependent on the solar distance from the Galactic centre:

\setlength{\mathindent}{0pt}
\begin{eqnarray}
D_{s}(R)=n_{s}~exp[10.093(1-R/R_{0})^{1/4}]/(R/R_{0})^{7/8}.
\end{eqnarray}

Since star counts at high Galactic latitudes are not strongly related to the radial distribution, they are well suited to study the vertical distribution of the Galaxy. Hence, we adopted the scalelength of the thin and thick discs, 2.4 and 3.5 kpc respectively, of Juri\'c et al. (\cite{J05}) based on $\sim$ 48 million stars, and we estimated the space densities for three components of the Galaxy at the solar distance, the scaleheights of the thin and thick discs, and the axial ratio of the halo by comparison of the space densities with the density laws in perpendicular direction. This is the procedure used by many researchers such as Phleps et al. (\cite{P00}), Siegel et al. (\cite{S02}), Du et al. (\cite{Du03}), KBH (\cite{KBH04}), Bilir, Karaali \& Tun\c{c}el (\cite{BKT05}), Phleps et al. (\cite{P05}), BKG (\cite{BKG06}) and Du et al. (\cite{Du06}).     	

\section{Metallicity calibration and absolute magnitude determination}

The metallicities of the sample stars with $0.12 <(g-r)_{0}\leq 0.95$, which corresponds to $0.3<(B-V)_{0}\leq1.1$, were evaluated using the following calibration (KBT):

\setlength{\mathindent}{0pt}
\begin{eqnarray}
[M/H] = 0.10-3.54\delta_{0.43}-39.63\delta^{2}_{0.43}+63.51\delta^{3}_{0.43},
\end{eqnarray}
where $\delta_{0.43}$ is the normalized UV-excess in {\em SDSS\/} photometry. This parameter's range $0\leq\delta_{0.43}\leq0.33$ covers the metallicity interval $-3<[M/H]\leq0.2$ dex.

Absolute magnitudes have been determined using the procedure described by KBT. They give the absolute magnitude offset from the Hyades main sequence as a function of both $(g-r)_{0}$ colour and $\delta_{0.43}$ UV-excess, as follows:

\setlength{\mathindent}{0pt}
\begin{eqnarray}
\Delta M^{H}_{g} = 
c_{3}\delta^{3}_{0.43}+c_{2}\delta^{2}_{0.43}+c_{1}\delta_{0.43}+c_{0},
\end{eqnarray}
where the coefficients $c_{i}$ (i=0, 1, 2, 3) are functions of $(g-r)_{0}$ colour (Table 2) and are adopted from the work of KBT. $\Delta M^{H}_{g}$ is defined as the difference in absolute magnitudes of a program star and a Hyades star of the same $(g-r)_{0}$ colour:

\setlength{\mathindent}{0pt}
\begin{eqnarray}
\Delta M^{H}_{g} = M^{*}_{g}-M^{H}_{g},
\end{eqnarray}
and the absolute magnitude for a Hyades star can be evaluated from the (Hyades) sequence, normalized by KBT:

\setlength{\mathindent}{0pt}
\begin{eqnarray}
M^{H}_{g} = -2.0987(g-r)^{2}-0.0008(u-g)^{2}\nonumber\\
                +0.0842(g-r)(u-g)+7.7557(g-r)\nonumber\\
                -0.1556(u-g)+1.9714.
\end{eqnarray}
The KBT calibration is available for $4<M(g)\leq8$ absolute magnitude interval where luminosity function is almost flat in the solar neighbourhood (Jahreiss \& Wielen \cite{JW97}). 

In a conical magnitude-limited volume, the distance to which intrinsically bright stars are visible is larger than the distance to which intrinsically faint stars are visible. The effect of this is that brighter stars are statistically overrepresented and the derived absolute magnitudes are too faint. This effect, known as Malmquist bias (Malmquist \cite{M20}), was formalized into the general formula,
\setlength{\mathindent}{0pt}
\begin{equation}
M(g)=M_{0}-\sigma^{2}{d\log A(g) \over dg},
\end{equation}
where $M(g)$ is the assumed absolute magnitude, $M_{0}$ is the absolute magnitude calculated for any star using KBT calibration, $\sigma$ is the dispersion of the KBT calibration, and $A(g)$ is the differential counts evaluated at the apparent magnitude $g_{0}$ of any star. The dispersion in absolute magnitude calibration is around 0.25 mag, that produces a correction due to the Malmquist bias of less than 0.07 mag. The correction of the Malmquist bias was applied to {\em SDSS} photometric data. Combination of the absolute magnitude $M(g)$ and the apparent magnitude $g_{0}$ of a star gives its distance $r$ relative to the Sun, i.e.,

\setlength{\mathindent}{0pt}
\begin{equation}
[g-M(g)]_{0}=5\log r-5.
\end{equation}

Logarithmic space densities $D^{*}=\log D+10$ have been calculated for the combination of three population components (thin and thick discs and halo), where $D=N/ \Delta V_{1,2}$; $\Delta V_{1,2}=(\pi/180)^{2}(\sq/3)(r_{2}^{3}-r_{1}^{3})$; $\sq$ denotes the size of the field (60 deg$^{2}$); $r_{1}$ and $r_{2}$ denote the lower and upper limiting distances of the volume $\Delta V_{1,2}$; $N$ is the number of stars per unit absolute magnitude; $r^{*}=[(r^{3}_{1}+r^{3}_{2})/2]^{1/3}$ is the centroid distance of the volume $\Delta V_{1,2}$; and $z^{*}=r^{*}\sin(b)$, $b$ being the Galactic latitude of the field center. The limiting distance of completeness is calculated from the following equation:

\setlength{\mathindent}{0pt}
\begin{eqnarray}
[g-M(g)]_{0} =  5 \log r_{l} - 5,\\
z_{l}=r_{l} \sin (b)	
\end{eqnarray}
where $g_{0}$ is the limiting apparent magnitude (15 and 22, for the bright and faint stars, respectively), $r_{l}$ is the limiting distance of completeness, and $M(g)$ is the corresponding absolute magnitude defining the interval $(M_{1}, M_{2}]$.

\section{Estimation of the Galactic model parameters}

A $\chi^{2}$ method was employed to fit the analytical density laws given in Section 3 for combined all three population, i.e. thin and thick discs and halo, to the observational based space densities with the additional constraint of producing local densities consistent with those derived from Hipparcos (Jahreiss \& Wielen \cite{JW97}), a procedure applied in our previous papers (KBH, BKG). All error estimates were made by changing Galactic model parameters until an increase or decrease by 1 in $\chi^{2}$ was achieved (Phleps et al. \cite{P00}). The comparison is shown in Fig. 3, and the resulting Galactic model parameters are given in Table 3 for two fields. Comparison of the Galactic model parameters for two fields give interesting results: the scaleheight of the thin disc for two fields, 206($\pm$9) pc (for north) and 198($\pm$9) pc (for south), are rather close to each other. However this is not the case for the thick disc, $H$=493($\pm$12) and 579($\pm$15) pc for the north and south fields, respectively. According to Narayan \& Jog (\cite{N02}) and Momany et al. (\cite{M06}) the difference between the scaleheights is due to the warp and flare. The space density of the thin disc at the solar distance is almost the same for two fields, but there is a considerable difference between the space densities of the thick discs, i.e. $\sim$16 and $\sim$10 per cent relative the space density of the thin disc, for the north and south fields respectively. For the Galactic model parameters of the halo, one can say that there is a good agreement between the north and south ones. Especially the axis ratio, $\kappa=0.45$ for both fields, is consistent with the values of Larsen (\cite{L96}, $\kappa=0.48$), Chen et al. (\cite{C01}, $\kappa=0.55$), and with the more recent one of Juri\'c et al. (\cite{J05}, $\kappa=0.50$).

\section {Metallicity}

We estimated the metal abundances of unevolved G type stars with $5<M(g)\leq6$ by means of eq. (8), and we looked for a probable metallicity gradient in the vertical direction of the Galaxy. These are long lived stars which carry the heritage of the original chemical composition of the Galaxy, thus providing important information about the formation history of our Milky Way Galaxy. The metallicity distributions for 12 distance intervals are given in Fig. 4 and Fig. 5 for the north and south fields, respectively. In the panels from (a) to (e) and from (i) to (l), in both figures, one can notice a peak shifting towards low metallicies when one goes from short to large distances ($r$) relative to the Sun. Whereas there is a flat distribution for the panels (f), (g) and (h) which cover the distance interval $4<r\leq7$ kpc or $3.2<z\leq4.6$ kpc, $z$ being the distance from the Galactic plane. We adopted the abscissa of the peaks as the metal abundance of the corresponding distribution whereas for the flat distributions we used the medians. The metal abundances for 12 distance intervals mentioned above for north and south fields are given in Table 4. The number of stars, N, centroid distance, $r^{*}$, and the corresponding distance in the vertical direction, $z^{*}$, are also given in the same table.            
             
The metallicity versus $z^{*}$ diagrams for two fields are given in Fig. 6. Three substructures are conspicuous in this figure, i.e. the smooth tendencies for the intervals $z^{*}<3$ kpc and $5<z^{*}<10$ kpc, and the steep one for the interval $3<z^{*}<5$ kpc. The tendencies in the first interval correspond to low metallicity gradients, $d[M/H]/dz=-0.16(\pm0.02)$, and $-0.22(\pm0.02)$ dex kpc$^{-1}$ for the north and south fields, respectively. The interval $z^{*}<3$ kpc involves the thin disc which is formed by dissipative collapse of the corresponding part of the Galaxy. Hence, the cited metallicity gradients are (partly) due to the formation procedure  of the thin disc. The metallicity gradients detected for two fields for the interval $5<z^{*}<10$ kpc are almost zero, $d[M/H]/dz=-0.036(\pm0.001)$ and $-0.034(\pm0.001)$ dex kpc$^{-1}$ for the north and south fields, again consistent with the formation scenarios of the halo (see Section 7 for detail). The metallicity gradients for two fields for the interval $3<z^{*}<5$ kpc are rather different from each other, i.e. $d[M/H]/dz=-0.36(\pm0.12)$ and $-0.18(\pm0.01)$ dex kpc$^{-1}$, for the north and south fields respectively. Also, the first value is the highest gradient detected in this work. The interval $3<z^{*}<5$ kpc covers both thick disc and halo stars. Hence, the metallicity gradients cited for this interval probably correspond to the transition from thick disc to halo. The combination of the data in Table 4, for two fields individually, give metallicity gradients almost equal to each other for the whole Galaxy, $d[M/H]/dz=-0.29(\pm0.04)$ and $-0.30(\pm0.04)$ dex kpc$^{-1}$ for the north and south fields respectively.  

\section {Conclusion}

We estimated Galactic model parameters for two fields in the anti-centre direction of the Galaxy ($l=180^{o}$) and symmetric relative to the Galactic plane, $b=+45^{o}$ (north field) and $b=-45^{o}$ (south field). The size of the fields are large, avoiding any fluctuation in the values of the model parameters. We adopted the scalelengths for thin and thick discs from Juri\'c et al. (\cite{J05}), $h_{1}=2.4$ and $h_{2}=3.5$ kpc respectively, determined by means of $\sim$48 million stars, and we estimated the space densities at the solar distance for three Galactic components, scaleheights of thin and thick discs, and the axis ratio for the halo. 

No conspicuous differences could be detected between the corresponding Galactic model parameters for the thin discs of two fields. However this is not the case for the thick discs. Actually, the scaleheights are 493 and 579 pc, and the space densities at the solar distance normalized to the thin disc are 16 and 10 per cent for the north and south fields respectively. According to Narayan \& Jog (\cite{N02}), and Momany et al. (\cite{M06}), these differences originate from warps and flares. For halo, the axis ratio of two fields are exactly the same, $\kappa=0.45$, and the space densities at the solar distance are rather close to each other.

Comparison of our model parameters with the ones of Juri\'c et al. (\cite{J05}) from which the scalelengths of the thin and thick discs were adopted shows both agreement and differences between two sets of data. For example, Juri\'c et al. (\cite{J05}) give $\kappa \sim0.5$ for the axis ratio of the halo which is almost equal to the value cited in our work. However, their scaleheights of the thin and thick discs are different than ours, especially the scaleheights of the thick disc for both fields are less than the half of their scaleheight, i.e. $H_{2}\sim1200$ pc. The scaleheight of the thick disc cited by Juri\'c et al. (\cite{J05}) reminded us the original value of Gilmore \& Reid (\cite{GR83}), $H=$1-1.4 kpc, however it is rather different than  the updated value (cf. Chen et al. \cite{C01}). Another conspicuous difference is between the normalized space densities of the thick disc at the solar distance. Juri\'c et al. (\cite{J05}) give a local thick-to-thin disc normalization 4 per cent which is 0.25 and 0.40 times of the values cited in our work for the north and south fields respectively. Our local densities for the thick disc are close to the updated ones (cf. Chen et al. \cite{C01}) whereas the one of Juri\'c et al. (\cite{J05}) is close to the original one, $\sim2$ per cent. Despite these differences, the Galactic model parameters cited in our work are consistent with the corresponding ones appeared in the literature (see Table 1) with two exceptions, i.e. the local space density normalization for the thick disc for the north field, 16 per cent, and the scaleheight of the thick disc for the north field, 493 pc.
 
Three substructures can be revealed for the metallicity distribution of stars with $5<M(g)\leq6$: 
\begin{itemize}
\item[(1)]
For $z^{*}<3$ kpc the metallicity gradients are smooth and close to each other for two fields, $d[M/H]/dz=-0.16(\pm0.02)$, and $-0.22(\pm0.02)$ dex kpc$^{-1}$ for the north and south fields respectively. 
\item[(2)]
For the interval with stars at large distances, i.e. $5<z^{*}<10$ kpc, the metallicity gradient is $d[M/H]/dz \sim-0.03(\pm0.004)$ dex kpc$^{-1}$ for two fields.
\item[(3)]
For the intermediate distance interval $3<z^{*}<5$ kpc, the metallicity gradients for two fields are rather different from each other, $d[M/H]/dz= -0.36(\pm0.12)$ and $-0.18(\pm0.01)$ dex kpc$^{-1}$ for the north and south fields respectively. Also, the tendency of the metallicity for stars in the north field is rather steep. The combination of the metallicity distributions in these substructures gives a metallicity gradient of $d[M/H]/dz \sim -0.30\\ (\pm0.04)$ dex kpc$^{-1}$ for two fields. 
\end{itemize}

The result cited in (1) is consistent with the formation scenario of the thin disc, i.e. this component of the Galaxy was formed by dissipative collapse. The smooth gradient is due to the contamination of the thick disc for which no metallicity gradient is expected. The interval $5<z^{*}<10$ kpc covers predominantly the halo stars, and the metallicity close to zero is the expected one. That is, halo is subject to some mergers or accreation of the objects formerly formed outside of our Galaxy. The most interesting result is for the interval $3<z^{*} <5$ kpc: the metallicity gradients for the north and south fields are different from each other. This interval is partly occupied by the thick disc stars. Hence, the metallicity gradient cited for this interval corresponds to the transition region from the thick disc to the halo. We remind the reader the differences between the scaleheights and local space densities of the north and south fields cited above. These findings encariage us to argue that the structure of the thick disc is Galactic latitude dependent. 

The metallicity gradients cited in our work are consistent with the corresponding ones appeared in the literature. For example, Trefzger et al. (\cite{T95}) estimated a metallicity gradient of $d[M/H]/dz=-0.23(\pm0.04)$ dex kpc$^{-1}$ for $z<4$ kpc, and Karaali et al. (\cite{K03}) give $d[M/H]/dz=-0.20$, -0.10, and –0.20 dex kpc$^{-1}$ for the intervals $z<5$, $5<z\leq8$, and $z\leq8$  kpc. For a more recent citation, we can give that of Du et al. (\cite{Du04}): $d[M/H]/dz=-0.37(\pm0.01)$, $-0.06(\pm0.01)$ and $-0.17(\pm0.04)$ dex kpc$^{-1}$ for $z<4$, $z>4$ and $z<15$ kpc.       

\section*{Acknowledgments}
This work was supported by the Research Fund of the University of Istanbul. Project number: BYP 705/07062005. We thank H. \c{C}akmak for 
preparing computer program for this study.

\begin{figure*}
\begin{center}
\includegraphics[angle=0, width=170mm, height=101.6mm]{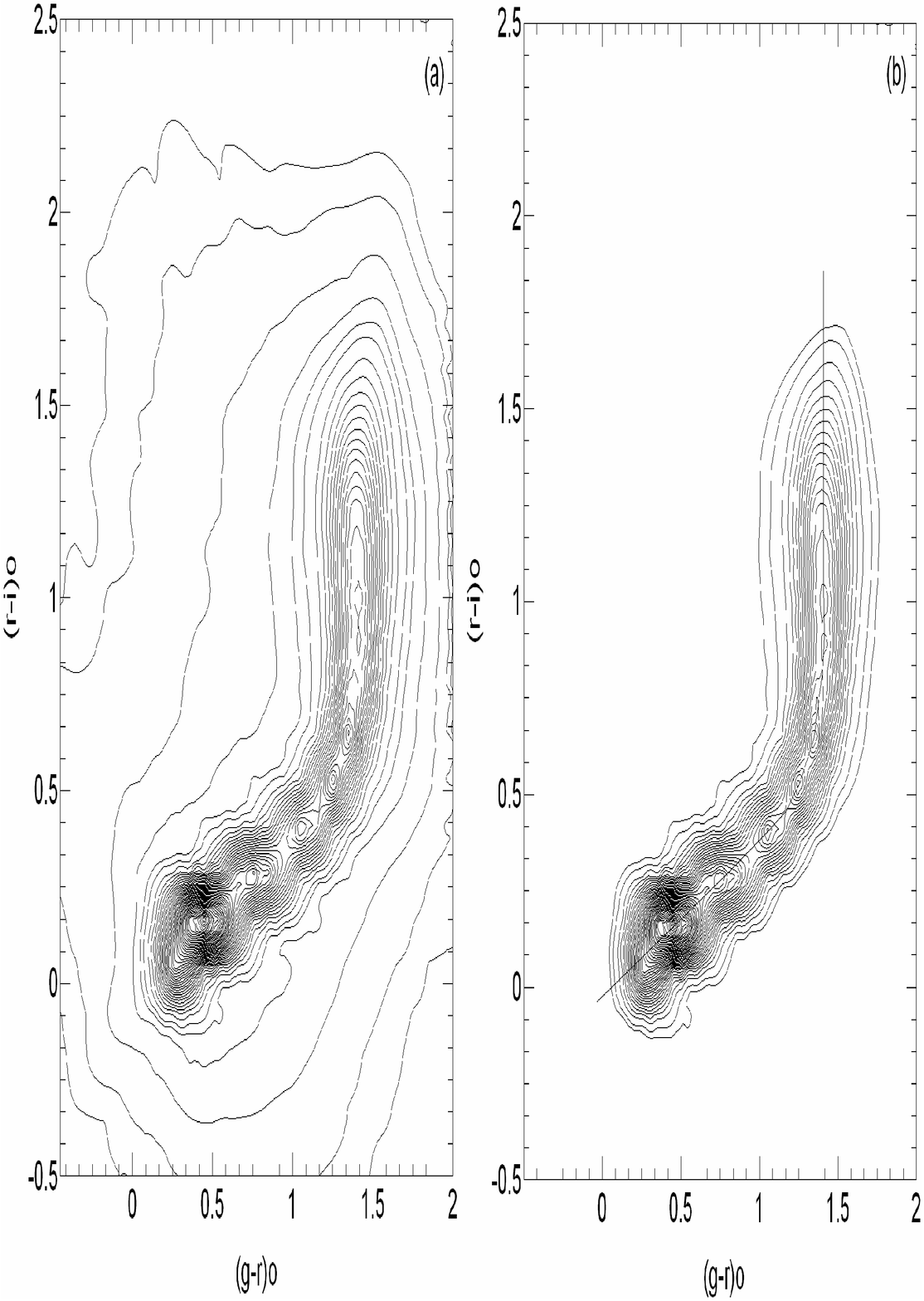}
\caption[] {The distribution of point sources (a) and stars (b) in our work in the $(g-r)_{o}-(r-i)_{0}$ two colour diagrams, shown by isodensity contours.}
\end{center}
\end{figure*}

\begin{figure*}
\begin{center}
\includegraphics[angle=0, width=170mm, height=97mm]{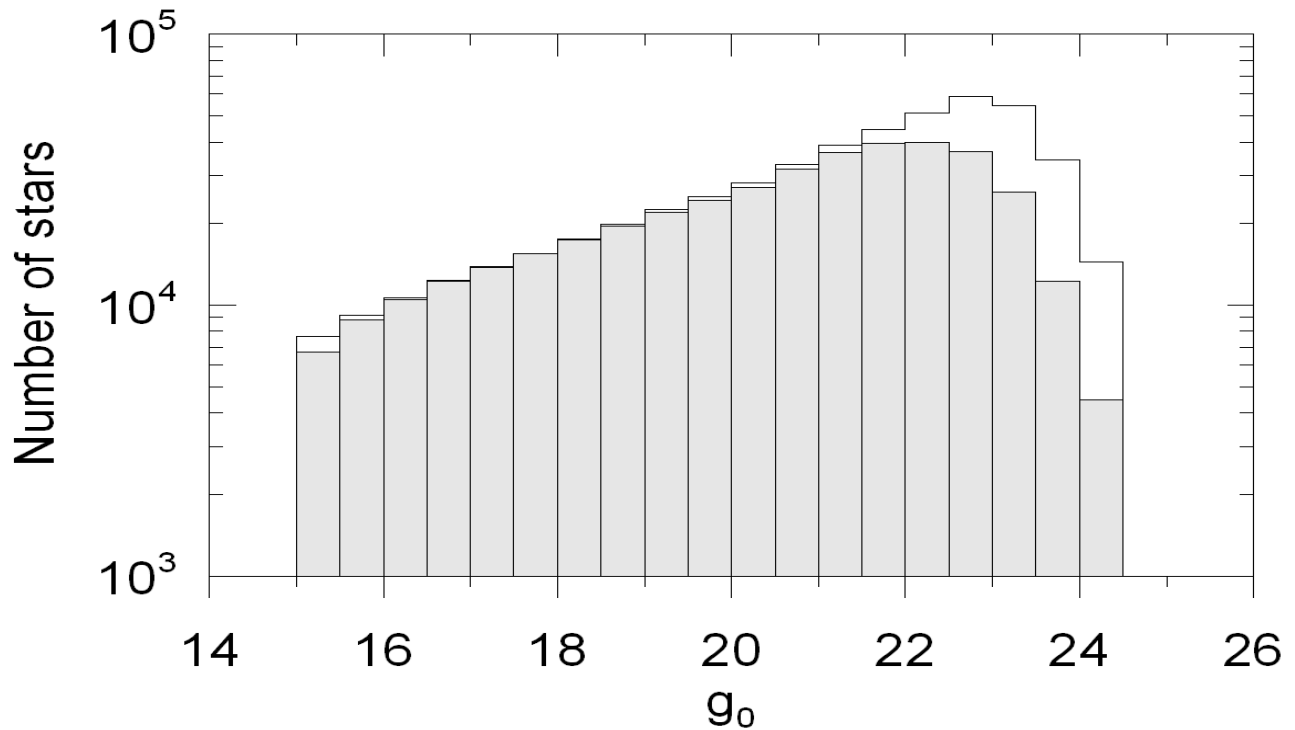}
\caption[] {Apparent magnitude histogram for point sources (white area) and for final stars sample (shaded area).}
\end{center}
\end{figure*}

\begin{figure*}
\begin{center}
\includegraphics[angle=0, width=170mm, height=174.6mm]{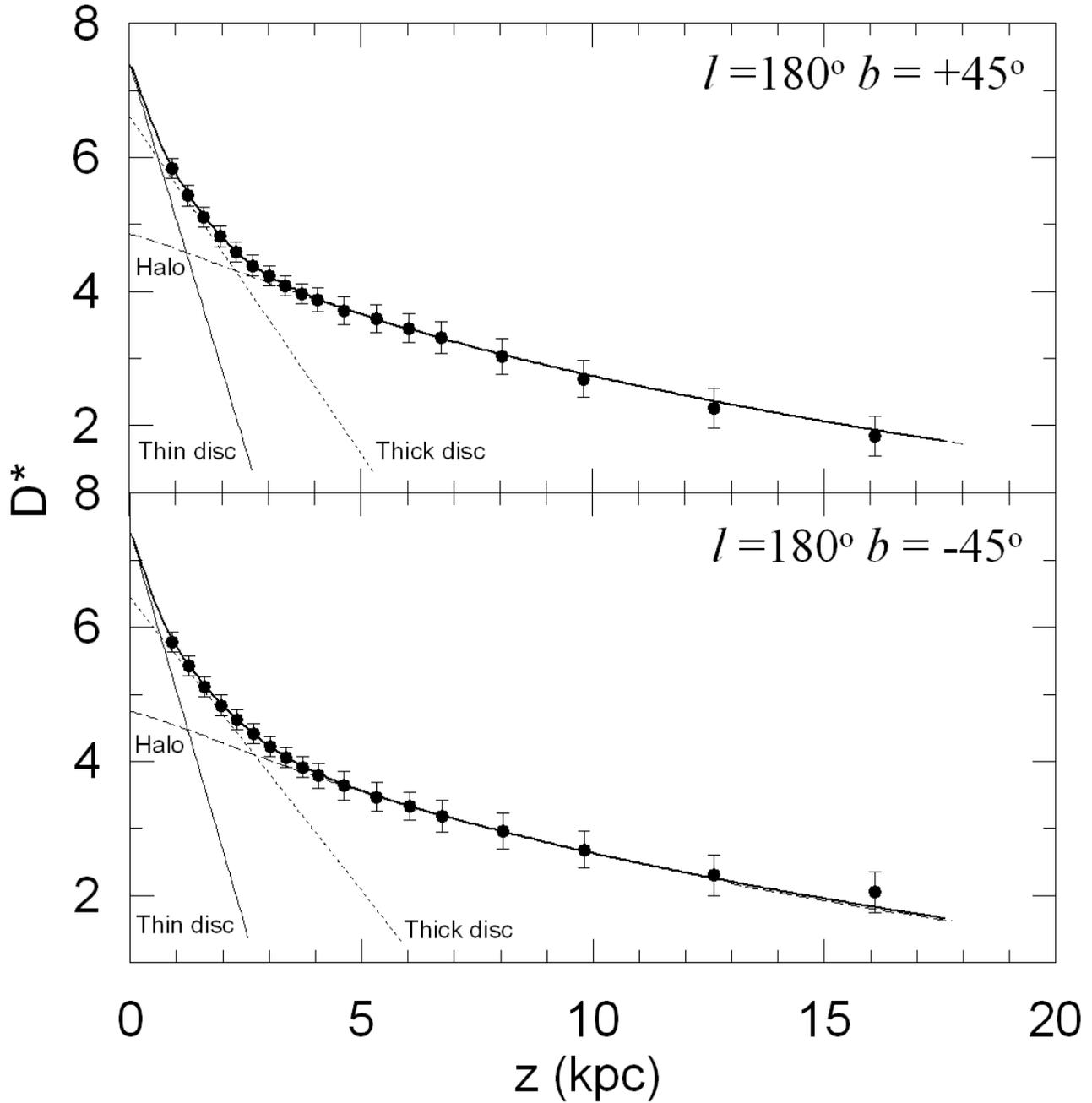}
\caption[] {Comparison of observed (symbols) and calculated (lines) space density functions combined for stars of all three population components and with absolute magnitudes $4<M(g)\leq8$ for two star fields.}
\end{center}
\end{figure*}

\begin{figure*}
\begin{center}
\includegraphics[angle=0, width=176.5mm, height=107.0mm]{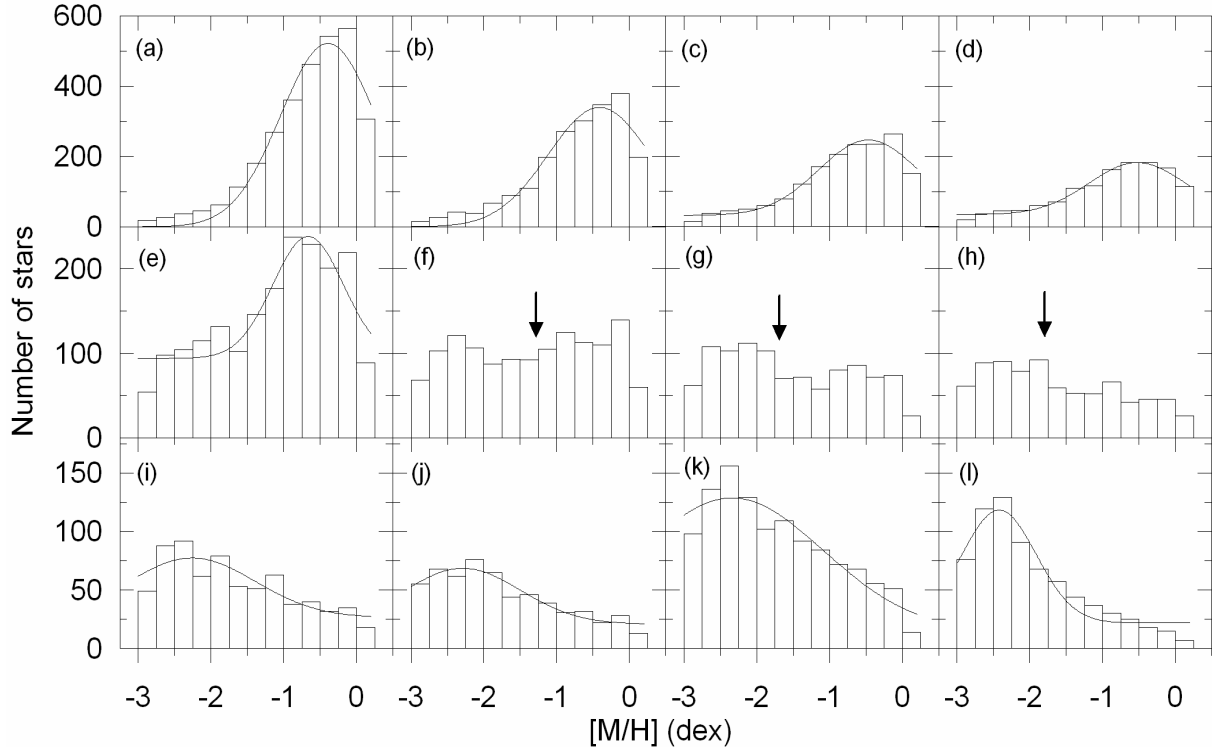}
\caption[] {Metallicity distributions for stars with $5<M(g)\leq 6$ absolute magnitude within the $r$-distance ranges (in kpc) for north field: (a) (0.5,1.5], (b) (1.5,2], (c) (2,2.5], (d) (2.5,3], (e) (3,4], (f) (4,5], (g) (5,6], (h) (6,7], (i) (7,8], (j) (8,9], (k) (9,12] and (l) (12,15]. The curves are Gaussian fits with mean metallicities and dispersion in Table 3. The arrow in the panels shows the median.}
\end{center}
\end{figure*}

\begin{figure*}
\begin{center}
\includegraphics[angle=0, width=176.5mm, height=107.0mm]{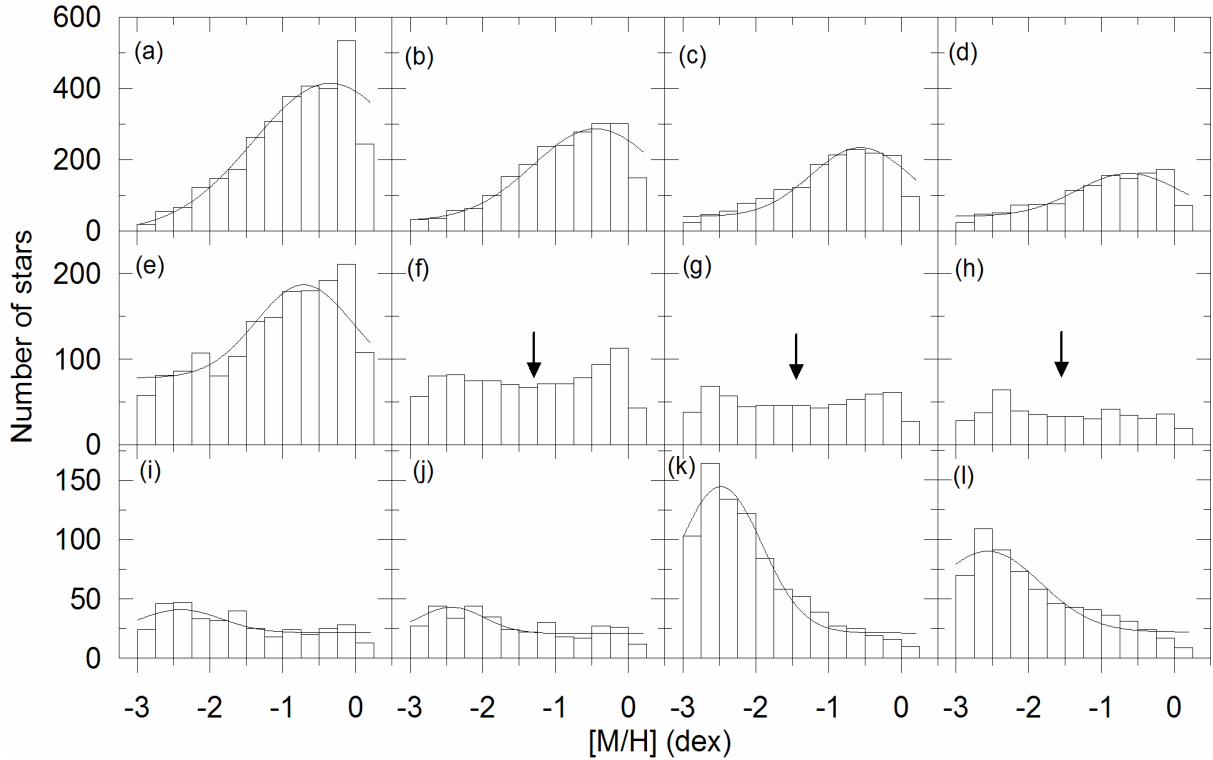}
\caption[] {Metallicity distributions for stars with $5<M(g)\leq 6$ absolute magnitude within the $r$-distance ranges (in kpc) for south field: Symbols are the same in Fig. 4.}
\end{center}
\end{figure*}

\begin{figure*}
\begin{center}
\includegraphics[angle=0, width=170mm, height=174.6mm]{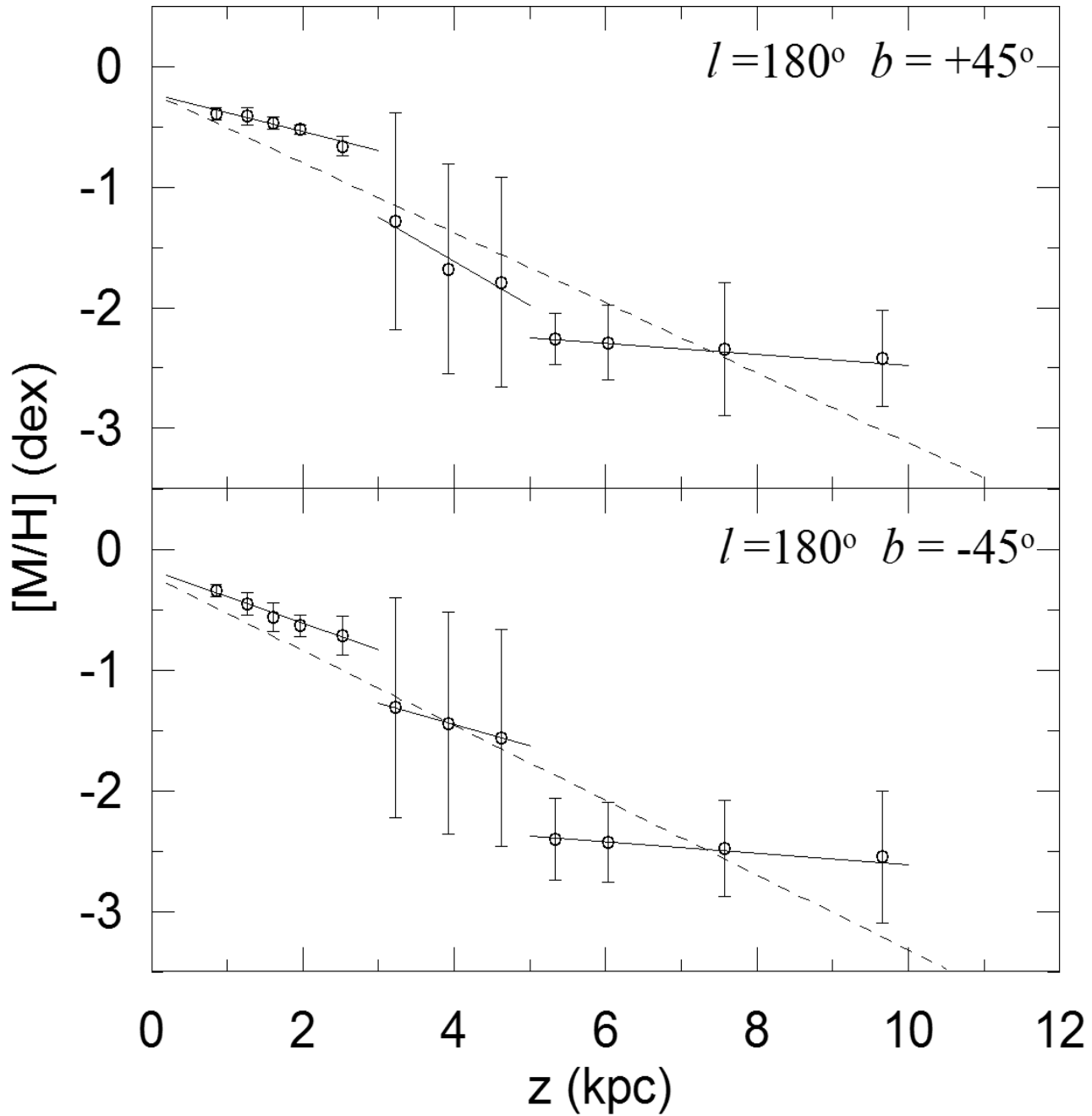}
\caption[] {Mean metal abundances versus $z^{*}$-distance for 12 $z^{*}$ intervals, showing the metallicity gradients $-0.29(\pm0.04)$ and $-0.30(\pm0.04)$ dex kpc$^{-1}$ up to 10 kpc for overall distribution, $-0.16(\pm0.02)$ and $-0.22(\pm0.02)$ dex kpc$^{-1}$ for $z^{*}<3$ kpc, $-0.36(\pm0.12)$ and $-0.18(\pm0.01)$ dex kpc$^{-1}$ for $3<z^{*}\leq5$ kpc, $-0.036(\pm0.001)$ and $-0.034(\pm0.001)$ dex kpc$^{-1}$ for $5<z^{*}\leq10$ kpc for north and south fields, respectively.}
\end{center}
\end{figure*}

\begin{table*}
\center
\caption{Previous Galactic models. Symbols: TN denotes the thin disc, TK  
denotes 
the thick disc, S denotes the spheroid (halo), $R_{e}$ is the effective radius and $\kappa=c/a$ is the axis ratio. The figures in the parentheses for Siegel et al. (\cite{S02}) are the corrected values for binarism. The asterisk denotes the power-law index replacing $R_{e}$.}
\begin{tabular}{lllllllll}
\hline
H (TN)& h (TN)& n (TK)& H (TK) & h (TK) & n (S) & $R_{e}$(S) & $\kappa$&  Reference \\
(pc) & (kpc) & &  (kpc) & (kpc) & & (kpc) & & \\
\hline
310-325 & --- & 0.0125-0.025 &  1.92-2.39 & --- & --- & --- & --- & Yoshii (\cite{Y82})\\
300  & --- & 0.02 & 1.45 & --- & 0.0020 & 3.0 & 0.85 & Gilmore \& Reid (\cite{GR83})\\
325  & --- & 0.02 & 1.3 &  --- & 0.0020 & 3.0 & 0.85 & Gilmore (\cite{G84})\\
280  & --- & 0.0028 &  1.9 & --- &0.0012 &--- & --- & Tritton \& Morton (\cite{TM84}\\
200-475 & --- & 0.016 &  1.18-2.21 & --- & 0.0016 & --- & 0.80 & Robin \& Cr\'{e}z\'{e} (\cite{R86})\\
300  & --- & 0.02 & 1.0 & --- & 0.0010 & --- & 0.85 & del Rio \& Fenkart (\cite{del87})\\
285  & --- & 0.015 &1.3-1.5 & --- & 0.0020 & 2.36 & flat & Fenkart et al. (\cite{F87})\\
325  & --- & 0.0224 & 0.95 & --- & 0.0010 & 2.9 & 0.90 & Yoshii et al. (\cite{Y87})\\
249  & --- & 0.041 & 1.0 & --- & 0.0020 & 3.0 & 0.85 & Kuijken \& Gilmore (\cite{KG89})\\
350  & 3.8 & 0.019 & 0.9 & 3.8 & 0.0011 & 2.7 & 0.84 & Yamagata \& Yoshii (\cite{YY92})\\
290  & --- & --- & 0.86 & --- & --- & 4.0 & --- & von Hippel \& Bothun (\cite{von93})\\
325  & --- & 0.0225 & 1.5 & --- & 0.0015 & 3.5 & 0.80 & Reid \& Majewski (\cite{RM93})\\
325  & 3.2 & 0.019 & 0.98 & 4.3 & 0.0024 & 3.3 & 0.48 & Larsen (\cite{L96})\\
250-270 & 2.5 & 0.056 & 0.76 & 2.8 & 0.0015 & 2.44-2.75* &  0.60-0.85 & Robin et al. (\cite{R96}); Robin et al. (\cite{R00})\\
290  &4.0 & 0.059 & 0.91 & 3.0 & 0.0005 & 2.69 & 0.84 & Buser, Rong \& Karaali (\cite{BRK98}, \cite{BRK99})\\
240  &2.5 & 0.061 & 0.79 & 2.8 & --- & --- & 0.60-0.85 & Ojha et al. (\cite{O99}) \\
330  &2.25& 0.065-0.13 &  0.58-0.75 & 3.5 & 0.0013 & --- & 0.55 & Chen et al. (\cite{C01}) \\
280  & 2-2.5&  0.06-0.10 & 0.7-1.0 & 3-4 & 0.0015 & --- & 0.50-0.70 & Siegel et al. (\cite{S02})\\
(350)& 2-2.5&  0.06-0.10 & (0.9-1.2) & 3-4 & 0.0015 & --- & 0.50-0.70 & Siegel et al. (\cite{S02})\\
320  & ---  & 0.07 & 0.64 & ---  & 0.0013 & --- & 0.58 & Du et al. (\cite{Du03}) \\
280  & 2.4  & 0.04 & 1.2 & 3.5 & --- & --- & 0.50 & Juri\'c et al. (\cite{J05}) \\
\hline
\end{tabular} 
\end{table*}

\begin{table*}
\center
\caption{Numerical values for the coefficients $c_{i}$ (i=0, 1, 2, 3) in eq. (9), adopted from the work of KBT (\cite{KBT05}).}
\begin{tabular}{crrrr}
\hline
$(g-r)_{o}$ &\multicolumn{1}{c}{$c_{3}$}&\multicolumn{1} {c}
{$c_{2}$} & \multicolumn{1} {c} {$c_{1}$}& \multicolumn{1}
{c} {$c_{0}$}\\
\hline
(0.12,0.22] &   -68.1210 &    26.2746 &     2.2277 &    -0.0177 \\
(0.22,0.32] &   -32.5618 &     6.1310 &     5.7587 &     0.0022 \\
(0.32,0.43] &     8.2789 &    -7.9259 &     6.9140 &     0.0134 \\
(0.43,0.53] &   -23.6455 &    -0.4971 &     6.4561 &     0.0153 \\
(0.53,0.64] &     0.2221 &    -5.9610 &     5.9316 &    -0.0144 \\
(0.64,0.74] &   -47.7038 &     0.1828 &     4.4258 &    -0.0203 \\
(0.74,0.85] &   -52.8605 &    12.0213 &     2.6025 &     0.0051 \\
(0.85,0.95] &   -15.6712 &     7.0498 &     1.6227 &    -0.0047 \\
\hline
\end{tabular}
\end{table*}

\begin{table*}
\center
\caption{Galactic model parameter values estimated for star fields. The columns give: logarithmic local space density $n^{*}$; scaleheight $H$; local space densities relative to thin disc, for thick disc $n_{2}/n_{1}$ and for halo $n_{3}/n_{1}$; axis ratio 
of halo $\kappa$.}
\begin{tabular}{ccccc}
\hline
Field & Parameter & Thin disc & Thick disc & Halo\\
\hline
North & $n^{*}$   & $7.40 \pm 0.09$ & $6.61 \pm 0.03$  & $4.90 \pm 0.06$\\
& $H$ (pc)  & $206\pm 9$ & $493\pm 12$ & $-$ \\
& $n/n_{1}($per cent$)$ & $-$ & $16.03 \pm 3$ & $0.31 \pm 0.03$ \\
& $\kappa$  & $-$ & $-$& $0.45 \pm 0.06$\\
\hline
South & $n^{*}$   & $7.43 \pm 0.09$ & $6.44 \pm 0.03$  & $4.79 \pm 0.09$\\
& $H$ (pc)  & $198 \pm 9$ & $579 \pm 15$ & $-$ \\
& $n/n_{1}($per cent$)$ & $-$ & $10.40 \pm 3$ & $0.23 \pm 0.03$\\
& $\kappa$  & $-$ & $-$& $0.45 \pm 0.09$\\
\hline
\end{tabular}
\end{table*}

\begin{table*}
\center
\caption{Metallicity distribution for 12 distance intervals, $r^{*}$ being the centroid distance, $z^{*}$ centroid distance to the Galactic plane, $[M/H]$ mean metallicity and N number of stars.}
\begin{tabular}{ccccccc}
\hline
& & & \multicolumn{2}{c}{North} & \multicolumn{2}{c}{South}\\
$r_{1}-r_{2}$& $r^{*}$& $z^{*}$ &      $[M/H]$   &   N &      $[M/H]$        &   N \\
  (kpc)      &  (kpc) &  (kpc)  &      (dex) &         &       (dex)         &     \\
\hline
   0.5-1.5 &  1.20 &  0.85 & $ -0.39 \pm 0.05$ &  2988 & $ -0.34 \pm 0.05$ &  3109 \\
   1.5-2.0 &  1.78 &  1.26 & $ -0.41 \pm 0.07$ &  2080 & $ -0.45 \pm 0.09$ &  2129 \\
   2.0-2.5 &  2.28 &  1.61 & $ -0.47 \pm 0.05$ &  1666 & $ -0.56 \pm 0.12$ &  1684 \\
   2.5-3.0 &  2.77 &  1.96 & $ -0.52 \pm 0.04$ &  1313 & $ -0.63 \pm 0.09$ &  1291 \\
   3.0-4.0 &  3.57 &  2.52 & $ -0.66 \pm 0.08$ &  1904 & $ -0.71 \pm 0.16$ &  1678 \\
   4.0-5.0 &  4.56 &  3.22 & $ -1.28 \pm 0.90$ &  1323 & $ -1.31 \pm 0.91$ &   975 \\
   5.0-6.0 &  5.54 &  3.92 & $ -1.68 \pm 0.87$ &  1026 & $ -1.44 \pm 0.92$ &   635 \\
   6.0-7.0 &  6.54 &  4.62 & $ -1.79 \pm 0.87$ &   801 & $ -2.56 \pm 0.90$ &   460 \\
   7.0-8.0 &  7.53 &  5.33 & $ -2.26 \pm 0.21$ &   700 & $ -2.40 \pm 0.34$ &   375 \\
   8.0-9.0 &  8.53 &  6.03 & $ -2.29 \pm 0.31$ &   581 & $ -2.43 \pm 0.33$ &   360 \\
  9.0-12.0 & 10.71 &  7.57 & $ -2.34 \pm 0.55$ &  1167 & $ -2.48 \pm 0.40$ &   853 \\
 12.0-15.0 & 13.66 &  9.66 & $ -2.42 \pm 0.40$ &   716 & $ -2.55 \pm 0.56$ &   660 \\
\hline
\end{tabular} 
\end{table*}

\end{document}